\documentclass[usegraphicx,usenatbib,twocolumn]{mn2e}
\usepackage{bm,latexsym,amsmath,amssymb,amsfonts}
\newcommand*{\C}{{\rm c}}
\newcommand*{\D}{{\rm d}}
\newcommand*{\TA}{{\rm ta}}
\begin{document}
\title[Cluster formation and the Sunyaev-Zel'dovich power spectrum in modified gravity]
{
Cluster formation and the Sunyaev-Zel'dovich power spectrum in modified gravity:
the case of a phenomenologically extended DGP model
}

\author[Kobayashi, T. and Tashiro, H.]
{Tsutomu Kobayashi$^1$ and Hiroyuki Tashiro$^2$\\
$^1$Department of Physics, Waseda University,
Okubo 3-4-1, Shinjuku, Tokyo 169-8555, Japan
\\
$^2$ Institut d'Astrophysique Spatiale (IAS), B\^atiment 121, F-91405, Orsay, France;\\
Universit\'e Paris-Sud 11 and CNRS (UMR 8617)
}


\maketitle

\begin{abstract}
We investigate the effect of modified gravity on cluster abundance
and the Sunyaev-Zel'dovich angular power spectrum.
Our modified gravity is based on a phenomenological extension of
the Dvali-Gabadadze-Porrati model which includes two free parameters
characterizing deviation from $\Lambda$CDM cosmology.
Assuming that Birkhoff's theorem gives a reasonable approximation,
we study the spherical collapse model of structure formation and
show that while the growth function changes to some extent,
modified gravity gives rise to no significant change in 
the linear density contrast at collapse time.
The growth function is enhanced in the so called normal branch, while 
in the ``self-accelerating'' branch it is suppressed.
The Sunyaev-Zel'dovich angular power spectrum is computed
in the normal branch, which allows us to
put observational constraints on the parameters of the modified gravity model
using small scale CMB observation data.
\end{abstract}



\section{Introduction}

General Relativity is surely the most successful theory of gravity
that passes accurate tests in the solar system and laboratories.
However,
current cosmological observations indicate the presence of
dark matter and dark energy;
a large fraction of the Universe is made of unknown components.
The mystery of the dark components is based on general relativity,
and hence it tells us that what we do not know may be the long distance behaviour
of gravity rather than the energy-momentum components in the Universe.
In this sense, cosmological observations
open up a new window to study the properties of gravity on large scales.

There are various alternative theories of gravity leading to
interesting cosmological consequences.
For example, the Dvali-Gabadadze-Porrati (DGP) model~\citep{DGP}
is one of the extra dimensional scenarios that can account for
cosmic acceleration without introducing dark energy.
The $f(R)$ theories, which modify the four-dimensional Einstein-Hilbert action explicitly,
also realise the accelerated expansion of the Universe~\citep{f(R)}.
MOND~\citep{MOND} and
its relativistic extension~\citep{BeMOND} explain galactic rotation curves without need for dark matter.
A more phenomenological way of changing gravity is to assume
Yukawa-like modification to a gravitational potential.
Such modification yields effectively a scale-dependent Newtonian constant,
and its effect on the evolution of large scale structure has been
investigated~\citep{Sealfon, Shirata1, Shirata2, Stabenau, Martino}.

One of the powerful ways for distinguishing modified gravity from
the $\Lambda$CDM model is to study the growth function because
the modified growth function would leave its footprints on
the cosmological large-scale structure.
In this context, many authors have studied the integrated
Sachs-Wolfe effect and weak lensing in modified gravity,
and obtained observational constraints on modified gravity models.
For instance,
\citet{Schmidt-2008} investigated
the effect of $f(R)$ gravity, the DGP model, and tensor-vector-scalar theory
on weak lensing,
and showed that for detecting signatures of modified gravity
the weak lensing observation is a better probe than the integrated Sachs-Wolfe effect measured 
via the galaxy-CMB cross-correlation.
\citet{Thomas-2008} put constraints on the DGP model
by using weak lensing data (CFHTLS-wide) combined with 
baryon acoustic oscillations
and supernovae data.
\citet{schmidt-lima-2009} have studied the statistical properties of
dark halos in $f(R)$ gravity by employing numerical simulations.


In this paper, we consider
a generalisation of the DGP model, adding a term $\pm H^{2\alpha}/r_c^{2(1-\alpha)}$
in the Friedmann equation, where $\alpha$ and $r_c$ are
the model parameters~\citep{DvaliTurner, Koyama^alpha, AGK, Khoury-N-body}.
The generalised model reduces
to the self-accelerating and normal branches of the DGP model
(with a cosmological constant)
by taking $\alpha\to1/2$,
and reproduces the $\Lambda$CDM model in the $\alpha\to 0$ limit.
\cite{AGK} studied cosmological perturbations in this modified gravity theory
under several assumptions, and discussed various observational consequences of the model.
\cite{Khoury-N-body} performed N-body simulations in the same model recently.
We shall work on a different approach in the present paper:
assuming that Birkhoff's theorem gives a reasonable approximation
(as implied by~\cite{Koyama^alpha}),
we study the spherical collapse model of structure formation
with the modified Friedmann equation.
Using the spherical collapse scenario,
we can compute the Sunyaev-Zel'dovich (SZ) angular power spectrum
through the Press-Schechter formalism~\citep{press-schechter-74}.
Since
the SZ angular power spectrum is sensitive to the distribution of dark halos,
it is a good probe of modified distribution of
dark halos in the generalised DGP model.

Before closing the introduction
we should note that in any case modified gravity must satisfy
solar system and laboratory tests, and some $f(R)$ models and
the original DGP model indeed have mechanisms to reproduce
ordinary gravity around the Sun and on the Earth.
This point is not clear in the present model and is beyond the scope of the paper.
We only remark that our Friedmann equation
reduces to the ordinary one at high densities.

This paper is organized as follows.
In the next section, we describe our modified gravity model in terms of the Friedmann equation
and explain its possible origin.
In Sec.~\ref{Sec:SF}, we review the spherical collapse model of structure formation
in modified gravity, and compute the linear growth function and
the linear density contrast at collapse time.
Then, in Sec.~\ref{Sec:SZ}, we study the effect of modified gravity on the
SZ angular power spectrum.
Observational constraints on the model parameters are discussed.
Finally, we conclude in Sec.~\ref{Sec:Conclusions}.




\section{The model}\label{Sec:Model}

The DGP braneworld was originally proposed
as a model for recovering 4D gravity on the brane
even in an infinitely large 5D Minkowski bulk~\citep{DGP}.
In the DGP braneworld, gravity is modified at long distances,
while standard 4D gravity is indeed reproduced at short distances
\citep[through a complicated nonlinear mechanism -- e.g.,][]{Tanaka, KoyamaSilva, Lue1, Lue2},
so that the model passes solar system and laboratory tests.
Probably the most intriguing consequence of the DGP braneworld
comes out in the modified Friedmann equation~\citep{Deff}:
\begin{eqnarray}
H^2 = \frac{8\pi G}{3}\rho \pm \frac{H }{r_c }.\label{DGP_Friedmann}
\end{eqnarray}
Here, $r_c$ is the crossover scale above which gravity looks 5D.
With the upper (plus) sign and $r_c\sim H_0^{-1}$ (the present Hubble horizon scale),
we have the ``self-accelerating'' solution which could be
the origin of the current cosmic acceleration.
In this paper, however, we do not assume that modified gravity is directly responsible for
the accelerated expansion.

The model we consider is
a phenomenological extension of the DGP braneworld
described by the modified Friedmann equation
\begin{eqnarray}
H^2 = \frac{8\pi G}{3}\rho + \frac{\Lambda}{3}\pm\frac{H^{2\alpha}}{r_c^{2(1-\alpha)}},
\label{modFriedmann}
\end{eqnarray}
where $0\le \alpha <1$.
This is similar to the model of~\cite{DvaliTurner},
but we allow for a different sign of the last term
and include the cosmological constant $\Lambda$ explicitly~\citep{AGK}.
With the upper (respectively lower) choice of
sign we use the terminology the self-accelerating (respectively normal) branch,
though it is not {\em self}-accelerating in the upper sign case.
Equation~(\ref{modFriedmann}) with $\alpha=1/2$ corresponds to
the DGP cosmology (with a cosmological constant or the tension on the brane),
while $\alpha=1$ can be absorbed into a redefinition of the gravitational constant $G$.
Expansion history of
$\Lambda$CDM cosmology is recovered in the limit $\alpha\to 0$.



\begin{figure}
  \begin{center}
    \includegraphics[keepaspectratio=true,height=50mm]{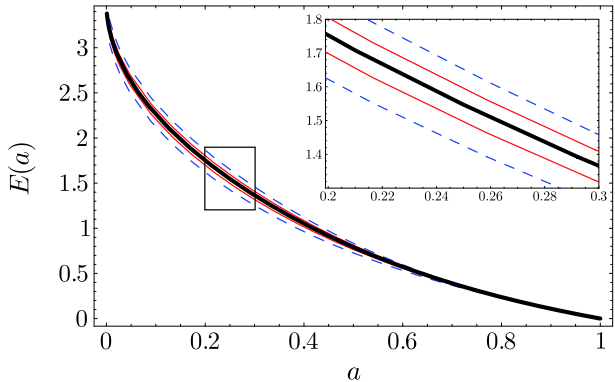}
  \end{center}
  \caption{Dimensionless physical distance. Parameters are (from top to bottom):
  (1) $\alpha=0.1$, $r_cH_0=0.4$;
  (2) $\alpha=0.01$, $r_cH_0=0.2$;
  (3) $\Lambda$CDM;
  (4) $\alpha=0.01$, $r_cH_0=0.2$;
  (5) $\alpha=0.1$, $r_cH_0=0.4$.
  The cases (1) and (2) are for the normal branch and
  the cases (4) and (5) are for the self-accelerating branch.
  In all the cases $\Omega_{{\rm m}0}=0.26$.
  }
  \label{fig:Ea.eps}
\end{figure}

The modified Friedmann equation~(\ref{modFriedmann})
can be recast in
\begin{eqnarray}
\left(\frac{H}{H_0}\right)^2=\frac{\Omega_{{\rm m}0}}{a^3}+\lambda
\pm (r_cH_0)^{-2(1-\alpha)}
\left(\frac{H}{H_0}\right)^{2\alpha},
\end{eqnarray}
where
\begin{eqnarray}
\lambda:=1-\Omega_{{\rm m}0}
\mp (r_cH_0)^{-2(1-\alpha)},
\end{eqnarray}
$\Omega_{{\rm m}0}:=8\pi G\rho_0/(3H_0^2)$,
and the present scale factor is chosen to be $a_0=1$.
If $\Lambda (\propto\lambda)= 0$ and the accelerated expansion were supported by modified gravity
in the self-accelerating branch,
successful cosmology would require $r_{c}\sim H^{-1}_{0}$.
In the present case, however, $\Lambda$ is responsible for the accelerated expansion
as in conventional cosmology, and therefore
we are in principle allowed to take $r_{c}<H^{-1}_{0}$ (e.g., $r_{c}\sim 0.1 H^{-1}_{0}$).\footnote{The
normal branch does not admit acceleration without $\Lambda$ from the beginning.}
In particular, for sufficiently small $\alpha$, expansion history is
very close to that in standard $\Lambda$CDM cosmology even
if we take relatively small $r_c$~\citep{AGK}.
Indeed, for $\alpha=0.1$, $r_cH_0=0.4$, and $\Omega_{{\rm m}0}=0.26$,
the dimensionless physical distance
\begin{eqnarray}
E(a):=H_0\int^1_{a}\frac{\D a}{a^2H(a)},
\label{distance}
\end{eqnarray}
differs from the corresponding $\Lambda$CDM result by less than one percent (Fig.~\ref{fig:Ea.eps}).

Unfortunately, we do not have concrete higher dimensional models
that account for Eq.~(\ref{modFriedmann}).
One possibility is that
such modification could be derived from a higher codimension DGP model~\citep{AGK},
as explained below.
Let us consider a graviton propagator which is proportional to
\begin{eqnarray}
\frac{1}{k^{2}+ k^{2\gamma}/r_{c}^{2(1-\gamma)}}.
\end{eqnarray}
This follows from a phenomenological model of modified gravity proposed in~\cite{Dvali} and is
a power-law generalisation of the graviton propagator in
the DGP braneworld.
Since we are interested in long distance modification of gravity, we assume that $\gamma <1$.
The unitarity constraint requires $\gamma\ge 0$~\citep{Dvali}.
It can be seen that
$\gamma=1/2$ reproduces the original DGP model and
$\gamma = 1$ can be absorbed into a redefinition of $G$.
From this observation, we may simply identify $\gamma = \alpha$.
The propagator with $\gamma\ll 1$ has some connection to
the so called cascading DGP braneworld~\citep{derham,  cascade1},\footnote{
The codimension-two case in fact gives the propagator $\sim\ln k$ for small $k$.}
and so
the Friedmann equation with $\alpha \ll 1$ might be realised
in such higher codimension models (see also \citealt{KK, TK}).
However, we would like to stress that
detailed analysis of higher codimension DGP models has yet to be undertaken
and no braneworld models have been known so far that lead to Eq.~(\ref{modFriedmann}).
Therefore, we shall view Eq.~(\ref{modFriedmann}) as a phenomenological starting point
of our modified gravity.

\section{Structure formation in modified gravity}\label{Sec:SF}

In the original DGP case, we have the covariant gravitational field equations
that not only lead to the Friedmann equation~(\ref{DGP_Friedmann}) but also
govern the behaviour of cosmological perturbations.
As we do not know a complete set of field equations
that underlies our phenomenologically extended model,
we must {\em assume} something about
the dynamics of cosmic inhomogeneities in the present case.
\cite{Koyama^alpha} clarified this issue by constructing
simple covariant gravitational equations which give
essentially the same Friedmann equation as Eq.~(\ref{modFriedmann}).
The gravitational equations contain an additional term called $E_{\mu\nu}$ in order to
satisfy the Bianchi identity, which hinders to get closed form equations.
In the original DGP braneworld, the evolution of the $E_{\mu\nu}$ term follows from
the full 5D Einstein equations. In the absence of underlying theories for general $\alpha$,
one has to assume the structure of $E_{\mu\nu}$, which then determines
the growth of structure.
\cite{Koyama^alpha} considered two possibilities:
(i) weak gravity is described by the scalar-tensor theory, as in the original DGP model;
(ii) the modified gravity model respects Birkhoff's theorem (at least approximately).
\cite{AGK} invoke the parameterized post-Friedmann framework~\citep{PPF} and hence
effectively take the first approach.
In this paper, we employ the second approach and
study nonlinear structure formation in modified gravity.
The assumed Birkhoff's theorem allows us to use
the modified Friedmann equation to track the nonlinear
dynamics in a simple way, rather than introduce the extra scalar degree of freedom explicitly.
The modified Friedmann equation reduces to the usual one in the high density regime,
and in this sense we implement the
nonlinear recovery of GR.
In the cases investigated by~\cite{Koyama^alpha},
the difference between the above two approach is small concerning
the linear growth of perturbations.\footnote{\cite{AGK}
focuses on the $\alpha\to 0$ limit, but in the PPF framework
cosmological perturbations are still sensitive to $r_c$.
On the other hand, the Birkhoff's theorem-based
approach relies essentially on the Friedmann equation,
and hence the effect of modified gravity vanishes in the $\alpha\to0$ limit.
Thus, the two approaches give different predictions at least in this limit.
}
We come back to this issue in Appendix.

In this section, we compute the linear growth function
and the linear density contrast for spherical collapse, which are
the key quantities for the Press-Schechter formalism~\citep{press-schechter-74}
to predict the number density of clusters.


\subsection{The growth equation}

\begin{figure}
  \begin{center}
    \includegraphics[keepaspectratio=true,height=50mm]{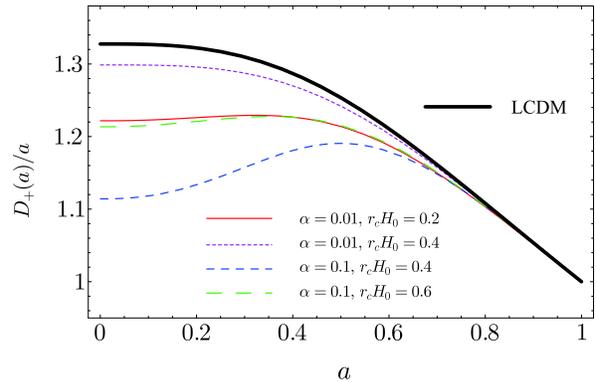}
  \end{center}
  \caption{The growth function divided by the scale factor for different parameters.
  Plots are for the normal branch and $\Omega_{{\rm m} 0}=0.26$.}%
  \label{fig:D1.eps}
\end{figure}
\begin{figure}
  \begin{center}
    \includegraphics[keepaspectratio=true,height=50mm]{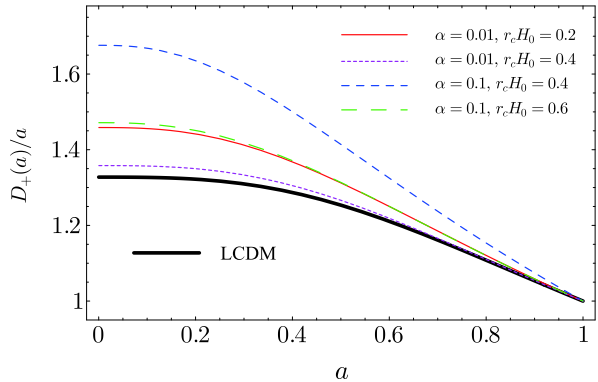}
  \end{center}
  \caption{The growth function divided by the scale factor for different parameters.
  Plots are for the self-accelerating branch and $\Omega_{{\rm m} 0}=0.26$.}%
  \label{fig:D2.eps}
\end{figure}

Following~\cite{BK},
we study a spherical overdensity with matter density
$\rho_\C=\rho_\C(t)$ and radius $R=R(t)$
in a background governed by the modified Friedmann equation~(\ref{modFriedmann}),
which can be recast in
\begin{eqnarray}
H^2=H_0^2 g(\xi),\quad \xi:=\frac{8\pi G\rho}{3H_0^2}.
\end{eqnarray}
Differentiation with respect to $t$ leads to
\begin{eqnarray}
\frac{\ddot a}{a}=H_0^2\left[g(\xi)-\frac{3}{2}\xi g'(\xi)\right],
\end{eqnarray}
where $\dot a:=\D a/\D t$ and $g': = \D g/\D\xi$.

Our central assumption is that {\em the modified gravity theory respects Birkhoff's theorem.}
The dynamics of $R(t)$ is then described by
\begin{eqnarray}
\frac{\ddot R}{R}=H_0^2\left[g(\xi_\C)-\frac{3}{2}\xi_\C g'(\xi_\C)\right],
\end{eqnarray}
with
\begin{eqnarray}
\xi_\C:= \frac{8\pi G\rho_\C}{3H_0^2}, \quad \rho_\C\propto r^{-3}.
\end{eqnarray}
We define the overdensity as
\begin{eqnarray}
\delta:=\frac{\rho_\C-\rho}{\rho}.
\end{eqnarray}
Upon linearisation the evolution equation for $\delta$ is given by
\begin{eqnarray}
\ddot\delta+2H\dot\delta=4\pi G\left[
g'(\xi)+3\xi g''(\xi)
\right]\rho\delta.\label{lin_delta}
\end{eqnarray}
The growth function $D_+(a)$ is defined by
$\delta (a, \mathbf{x})=D_+(a)\delta (a_0, \mathbf{x})$.
It follows from Eq.~(\ref{lin_delta}) that
\begin{eqnarray}
\frac{\D^2}{\D a^2}D_++\frac{f_1}{a}\frac{\D }{\D a}D_+=\frac{f_2}{a^2}D_+,
\end{eqnarray}
where
\begin{eqnarray}
f_1&=&3+\frac{\D\ln H}{\D\ln a},
\\
f_2&=&3\frac{\D\ln H}{\D\ln a}+ \left(\frac{\D\ln H}{\D\ln a}\right)^2
+\frac{a^2}{H}\frac{\D^2 H}{\D a^2}.
\end{eqnarray}
The boundary condition is given by $D_+(0)=0$ and $D_+(a_0)=1$.
Note that in standard cold dark matter dominated cosmology
we find $D_+(a) = a$.

Figures~\ref{fig:D1.eps} and~\ref{fig:D2.eps} show
typical behaviour of the growth function.
For example, 
the difference between background expansion history $E(a)$
in modified gravity with $(\alpha, r_cH_0)=(0.01,\,0.2)$ and that in the $\Lambda$CDM model
is $\lesssim 3\,\%$, for which
the growth functions differ by almost $10\,\%$ (both in the self-accelerating and normal branches).
In the case of the self-accelerating branch, the growth of perturbations is {\em suppressed}
compared to the $\Lambda$CDM model, in agreement with the result of~\cite{BK}.\footnote{Note that
the normalization of the growth function here is such that $D_+(a_0)=1$.
Hence, smaller (respectively larger) $D_+(a)$ implies
a smaller (respectively larger) amplitude of the perturbation at $a<1$
evolving into some fixed amplitude at $a=1$, i.e.,
the growth is enhanced (suppressed). 
}
This is because the effective dark energy term becomes dominant
earlier in modified gravity than in the $\Lambda$CDM model due to the $+H^{2\alpha}$ term.
In the normal branch, the effect of modified gravity works oppositely
and the growth of perturbations is {\em enhanced} compared to the $\Lambda$CDM model.
In the both cases, deviation from the $\Lambda$CDM result
is larger for larger $\alpha$ and smaller $r_c$, as expected.
Note here that the PPF approach
predicts qualitatively the same result:
structure is more evolved in the normal branch than in $\Lambda$CDM cosmology~\citep{AGK}.
Note also that qualitatively the same result, i.e.,
the suppressed (enhanced) growth function, is also found
in the self-accelerating (normal) branch of the original DGP model ($\alpha=1/2$)
by solving the five-dimensional Einstein equations~\citep{CKSS, Song}.

\subsection{Spherical collapse}

\begin{figure}
  \begin{center}
    \includegraphics[keepaspectratio=true,height=50mm]{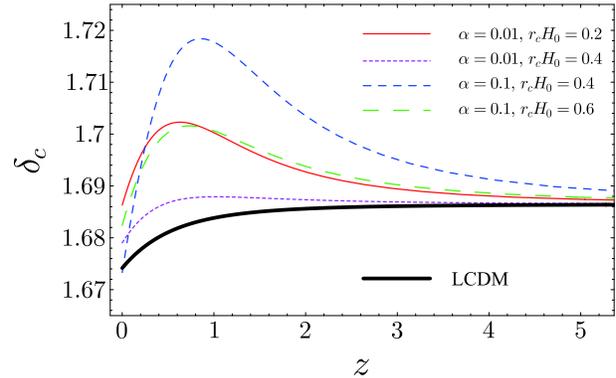}
  \end{center}
  \caption{Overdensity at collapse time for different parameters.
  Plots are for the normal branch and $\Omega_{{\rm m}0}=0.26$.}
  \label{fig:delta_c1.eps}
\end{figure}
\begin{figure}
  \begin{center}
    \includegraphics[keepaspectratio=true,height=50mm]{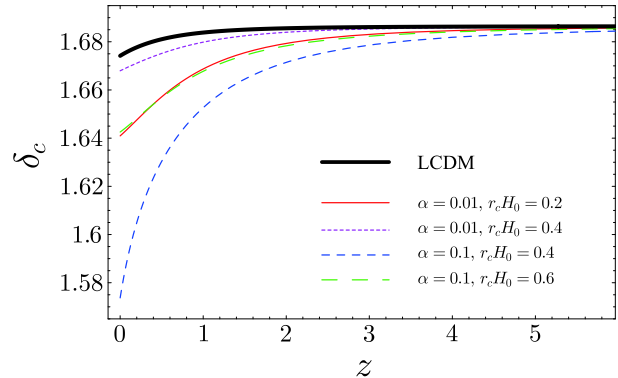}
  \end{center}
  \caption{Overdensity at collapse time for different parameters.
  Plots are for the self-accelerating branch and $\Omega_{{\rm m}0}=0.26$.}
  \label{fig:delta_c2.eps}
\end{figure}

In order to study spherical collapse,
it is convenient to use the quantities normalised by their values at turn-around
time~\citep{Wang, Mota, Bartelmann}.
First, we define the normalised scale factor and radius of the overdensity as follows:
\begin{eqnarray}
x:=a/a_\TA,\quad y:=R/R_\TA.
\end{eqnarray}
We also define the dimensionless time $\tau:=H_\TA t$, where $H_\TA:=H(a_\TA)$
is the Hubble rate at turn around.
Now the modified Friedmann equation can be written as
\begin{eqnarray}
&&\frac{H^2}{H_\TA^2}=
\left(\frac{\dot x}{x}\right)^2=\chi+1-\omega
\nonumber\\&&\qquad\qquad
\pm(H_\TA r_c)^{-2(1-\alpha)}\left[
\left(\frac{\dot x}{x}\right)^{2\alpha}-1
\right],
\label{ta_Friedmann}
\end{eqnarray}
where $\omega:=\Omega_{{\rm m}0}H_0^2/a_\TA^3 H_\TA^2$, $\chi:=\omega/x^3$,
and a dot here and hereafter denotes derivative with respect to $\tau$.
This equation can be rewritten as $(\dot x/x)^2=h(\chi)$, or, equivalently,
\begin{eqnarray}
\dot x=\sqrt{x^2 h\left(\frac{\omega}{x^3}\right)}.
\end{eqnarray}
Similarly to the previous calculation, one obtains
the another Friedmann equation that describes the evolution of the overdensity patch:
\begin{eqnarray}
\ddot y=y\left[
h\left(\frac{\zeta\omega}{y^3}\right)-
\frac{3}{2}\frac{\zeta\omega}{y^3}h'\left(\frac{\zeta\omega}{y^3}\right)
\right],\label{eqy}
\end{eqnarray}
where $h':=\D h/\D \chi$ and $\zeta:=(\rho_\C/\rho)|_{x=1}$.
The boundary condition $y|_{x=0}=0$, $\dot y|_{x=1}=0$, and $y|_{x=1}=1$
uniquely determines $\zeta$.
Equation~(\ref{eqy}) reduces to a first order differential equation by noticing that
\begin{eqnarray*}
\frac{\D}{\D\tau}\dot y^2 = \frac{\D}{\D\tau}\left[
y^2h\left(\frac{\zeta\omega}{y^3}\right)
\right].
\end{eqnarray*}
From this we obtain
\begin{eqnarray}
\dot y^2=y^2 h\left(\frac{\zeta\omega}{y^3}\right)
-h(\zeta\omega),
\label{eqyint}
\end{eqnarray}
where we fixed the integration constant by using the boundary condition
$\dot y=0$ at turn-around time ($y=1$).

Since the background dynamics at early times is the same as that of
the standard matter dominant universe, we may approximate
$h(\omega/x^3)\simeq \omega/x^3$ for $x\ll 1$. Thus, at early times we simply have
\begin{eqnarray}
\tau\simeq\frac{2}{3}\frac{x^{3/2}}{\sqrt{\omega}}.
\end{eqnarray}
Similarly, Eq.~(\ref{eqyint}) reduces to
\begin{eqnarray}
\D \tau\simeq \sqrt{\frac{y}{\zeta\omega}}\left[
1+ \frac{h(\zeta\omega)}{\zeta\omega}\frac{y}{2}\right]\D y
\quad\text{for}\quad y\ll 1,
\end{eqnarray}
leading to
\begin{eqnarray}
\tau\simeq\frac{2}{3}\frac{y^{3/2}}{\sqrt{\zeta\omega}}\left[
1+ \frac{3}{10}\frac{h(\zeta\omega)}{\zeta\omega}y\right].
\end{eqnarray}

The time evolution of the nonlinear overdensity, $\Delta:=\zeta x^3/y^3$, can be computed
at early times by using
\begin{eqnarray}
\Delta \simeq 1+\frac{3}{5}\frac{h(\zeta\omega)}{\zeta\omega}y.
\end{eqnarray}
The linear density contrast at collapse time, $\delta_c$, can be used to
relate the nonlinear overdensity $\Delta$ with the density that would
result from linear evolution and the same initial condition.
Using the growth function, we have
\begin{eqnarray}
\delta_\C:=\lim_{x\to 0}\frac{D_+(x_\C)}{D_+(x)}\left[\Delta(x)-1\right].
\end{eqnarray}

We numerically computed $\delta_c(z)$ for various model parameters, and
the results are illustrated in
Figs.~\ref{fig:delta_c1.eps} and~\ref{fig:delta_c2.eps}.
We find that modified gravity does not change $\delta_c$ much: for example,
$\delta_c$ in modified gravity with $(\alpha, r_cH_0)=(0.01, \,0.2)$
differs from the $\Lambda$CDM prediction
by 1 -- $2\,\%$ (both in the self-accelerating and normal branches).\footnote{It was reported that
in the original DGP model with $\Lambda=0$, the overdensity for spherical collapse
decreases by almost $20\,\%$ compared to conventional cosmology~\citep{BK}.
However, the first version of the preprint (i.e., arXiv:0711.3129, v1) includes
an error in the numerical calculations. We confirmed with the authors that
$\delta_c$ differs from the standard one only by a small amount
also in the case studied by~\cite{BK}.
}

\section{The Sunyaev-Zel'dovich power spectrum}\label{Sec:SZ}

In this section,
we investigate the effect of modified gravity on the SZ angular power spectrum.
Modification of gravity changes cluster number count through the modified
growth function and critical density contrast.
We concentrate on the case of the normal branch.
The SZ angular 
power spectrum could be amplified in this branch
because the growth function is 
enhanced relative to the $\Lambda$CDM model. 
The change in the critical density contrast is quite small,
and hence it will give rise to only a negligible effect.

Modified gravity might in general affect the halo profiles assumed in the following discussion.
This issue can be addressed by evaluating the Vainshtein radius, $r_*$, below which
Einstein gravity is recovered.
The Vainshtein radius in this particular model is
given by $r_* = [r_c^{4(1-\alpha)}r_{{\rm S}}]^{1/[1+4(1-\alpha)]}\simeq[r_c^{4}r_{{\rm S}}]^{1/5}$,
where $r_{{\rm S}}$ is the Schwarzschild radius of the source~\citep{Dvali}.
More explicitly, one has
$r_*\sim 10^2 [(r_cH_0)^4(M/M_\odot)]^{1/5}\,$kpc, and
for $r_cH_0\sim 0.01$ and $M\sim 10^{14}\, M_\odot$, $r_*\sim1\,$Mpc.
This estimate allows us to employ the halo profiles used in the context of standard gravity.

The computation of the SZ angular power spectrum is 
based on the halo formalism 
\citep{cole-kaiser-88, makino-suto-93, komatsu-kitayama-99, komatsu-seljak-02}.
Since the one-halo Poisson term dominates the halo-halo correlation term
on the scales we are interested in,  
we neglect the halo-halo correlation term.
The SZ angular power spectrum is then given by
\begin{eqnarray}
C_\ell ^{SZ}= g_\nu^2 \int_0^{z_{\rm max}}\! \D z \frac{\D V}{\D z}
\int_{M_{\rm min}}^{M_{\rm max}} \!\D M \frac{\D n(M,z)}{\D M}
\left|y_\ell(M,z)\right|^2,
\label{eq:cl^sz}
\end{eqnarray}
where $V(z)$ is the comoving volume at $z$ per steradian, 
$\D n(M,z)/\D M$ is the number density of clusters, 
$\tilde{y}_l(M,z)$ is the 2D Fourier transform of the projected 
Compton $y$-parameter,
and $g_\nu$ is the spectral function of the SZ effect, which is given by
\begin{eqnarray}
g_\nu = {x^2 e^x \over (e^{2x}-1)^2} \left[x \coth \left({1 \over 2}x \right)-4 \right],
\end{eqnarray}
where $x= h\nu /k_B T_{\gamma}$ with the Boltzmann constant $k_B$
and the CMB temperature $T_{\gamma}$. 

As we are considering the spherical collapse scenario for the formation of halos,
we utilise the Press-Schechter theory~\citep{press-schechter-74},
\begin{eqnarray}
{\D n(M,z) \over \D M} = \sqrt{2 \over \pi} { \rho \over M} 
\left[-{\delta_c \over \sigma(M,z)} {\partial \sigma \over\partial M} \right]
\exp \left[- { \delta_c ^2 \over 2 \sigma (M,z) }\right],
\label{eq:press-schechter}
\end{eqnarray}
where $\delta_c$ is the critical over density which is obtained in the previous section,
$\sigma (M,z)$ is the variance of the matter density field on the mass scale $M$.
The variance $\sigma$ is computed from the power spectrum
of linear matter density fluctuations with the top hat filter,
\begin{eqnarray}
\sigma ^2(M,z) = \int \D k \,k^2 P_{\rm m}(k,z) W(k R) ,
\label{eq:massdis}
\end{eqnarray}
where  $W(kR)$ is the top hat window function and 
$R$ is the scale which corresponds to $M  (=4\pi  \rho R^3/3)$. 
We calculate the linear power spectrum $P_{\rm m}(k,z)$ in the modified
gravity model by using the growth function obtained in the previous section 
and the initial condition for the curvature perturbation and the scalar spectral index,
$\Delta^2_{\cal R }= 2.41 \times 10^{-9}$,
$n=0.96$, given by WMAP \citep{wmap5-komatsu}.

The 2D Fourier transform of the projected Compton $y$-parameter
is given by
\begin{eqnarray}
y_\ell= {4\pi r_{\rm s} \over \ell_{\rm s}^2}
\int_0^\infty \D x\, x^2 y_{3d} (x)
{\sin(\ell x/\ell_{\rm s})  \over \ell x/\ell_{\rm s}},
\label{eq:yl}
\end{eqnarray}
where $y_{3d}$ is the radial profile of the Compton $y$-parameter,
\begin{eqnarray}
y_{3d} (x) =  {\sigma_{\rm T} \over m_e} ~ n_e(x) k_{\rm B}T_e (x),
\end{eqnarray}
with $\sigma_{\rm T}$ being the Thomson cross section 
and $m_e$ the electron mass.
$\ell_{\rm s}$ is the angular wavenumber corresponding to $r_{\rm s}$,
$\ell_{\rm s} = D_A / r_{\rm s}$, with $D_A$ being the angular diameter distance.

For the electron density profile $n_e$ and the temperature profile $T_e$, 
we use the result of \citet{komatsu-seljak-02},
which is based on the NFW dark matter density profile \citep{navarro-frenk}. 
The NFW dark matter density profile is given by 
\begin{eqnarray}
\rho_{\rm DM}(x) = \frac{\rho_{\rm s}}{x(1+x)^2},
\label{eq:NFWprofiles}
\end{eqnarray}
where $x := r/r_{\rm s}$ where $r_{\rm s}$ is a scale radius, and
$\rho_{\rm s}$ is a scale density. The scale radius
$r_{\rm s}$ is related to the virial radius by 
the concentration parameter $c$ as
\begin{eqnarray}
{r_{\rm s}(M,z)}={r_{\rm vir}(M,z) \over c(M,z)}.
\label{eq:scaleradius}
\end{eqnarray}
We use the concentration parameter of \citet{komatsu-seljak-02}, 
\begin{eqnarray}
c(M, z) \approx \frac{10}{1+z}\left[\frac{M}{M_*(0)}\right]^{-0.2},
\label{eq:concentrait}
\end{eqnarray}
where $M_*(0)$ is a solution to $\sigma(M,0)=\delta_c$ at the redshift $z=0$.

We adopt the spherical collapsed model to obtain the virial radius,
\begin{equation}
r_{\rm vir}(M,z) = {3 M \over 4 \pi \Delta_v(z) \rho (z) },
\end{equation}
where $\Delta_v(z)$ is the virialised overdensity at $z$.
Modified gravity changes the virialised overdensity,
which we compute 
following \citet{schmidt-lima-2009}.
For example, the modified virial overdensity 
at $z=0$ is $\Delta_v = 270$ for $(\alpha, r_cH_0)=(0.1, 0.3)$ and
$\Delta_v = 300$ for $(\alpha, r_cH_0)= (0.1, 0.4)$,
while $\Delta_v = 370$ in the $\Lambda$CDM model.
Thus, we find that the actual density in a virialised halo, $\Delta_v(z) \rho (z)$,
is lower in modified gravity
than in the $\Lambda$CDM model.


In order to obtain the profiles of the electron density and temperature,
\citet{komatsu-seljak-02}
assumed three things: (i) the electron gas is in hydrostatic
equilibrium in the dark matter potential; (ii) the electron gas density follows the
dark matter density in the outer part of the halo; (iii) the
equation of state of the electron gas is polytropic, $P_{e}\propto \rho_{e}^{\gamma}$, 
where $P_{e}$, $\rho_{e}$ and ${\gamma}$
are the electron gas pressure, the gas density and the polytropic index, respectively. 
Under 
these assumptions, 
the electron number density and temperature profiles 
are simply given by
\begin{eqnarray}
n_e &=& n_{e \rm c} F(x),
\\
T_e &=& T_{e \rm c} F^{\gamma -1}(x).
\end{eqnarray}
Here, $n_{e \rm c}$ and $T_{e \rm c}$ are
the electron density and temperature at the centre, respectively,
and the dimensionless profile $F(x)$ is written as
\begin{eqnarray}
F(x) = \left \{ 1-A \left[ 1- {\ln (1+c) \over x}\right]
\right \}^{1/(\gamma -1)},
\end{eqnarray}
where the coefficient $A$ is defined as
\begin{eqnarray}
A:=3\eta^{-1}_{\rm c}\frac{\gamma-1}{\gamma}
\left[\frac{\ln(1+c)}c-\frac1{1+c}\right]^{-1}.
\label{eq:Bcoefficient}  
\end{eqnarray}
The central electron density $n_{e \rm c}$ and the 
temperature $T_{e \rm c}$ are given by
\begin{eqnarray}
n_{e\rm c}
&=&3.01 
\left( {M \over 10^{14} M_\odot } \right)
\left( {r_{\rm vir} \over 1~ {\rm Mpc}  } \right)^{-3}
\nonumber \\
&& \!\!\!\!\!\!\!\!\! \times
\frac{c}{(1+c)^2}
\left[\ln(1+c)-\frac{c}{1+c}\right]^{-1}F^{-1}(c)
~{\rm cm}^{-3} ,
\label{eq:rhog0}
\\
T_{e\rm c}
&=&0.88  
~\eta_{\rm c} 
\left[ M / (10^{14} h^{-1} M_{\odot}) \over 
r_{\rm vir} / (1 h^{-1} {\rm Mpc}) \right]
~{\rm keV}.
\label{eq:rhog0}
\end{eqnarray}
For $\gamma$ and $\eta_{\rm c}$,
\citet{komatsu-seljak-02} provided the following useful fitting formulas:
\begin{eqnarray}
\gamma= 1.137 + 8.94\times 10^{-2}\ln(c/5) - 3.68\times 10^{-3}(c-5),
\label{eq:gammafitting}
\\
\eta_{\rm c}= 2.235 + 0.202(c-5) - 1.16\times 10^{-3}\left(c-5\right)^2.
\label{eq:etafitting}
\end{eqnarray}

\begin{figure}
  \begin{center}
  \includegraphics[keepaspectratio=true,height=60mm]{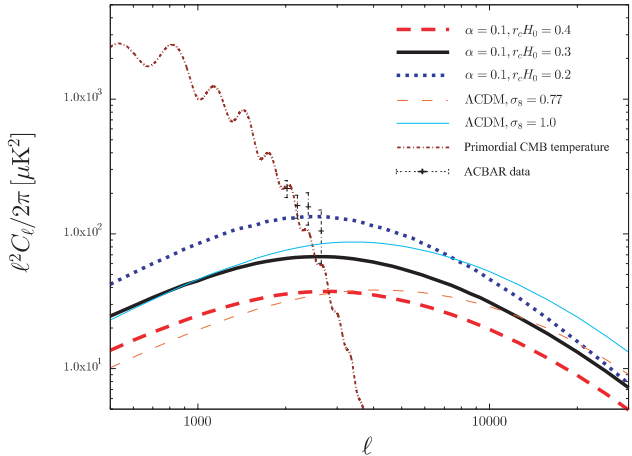}
  \end{center}
  \caption{
  SZ angular power spectra for different modified gravity parameters. 
  The dotted, solid, and dashed lines
  represent SZ power spectra for $(\alpha, r_cH_0)=(0.1, 0.2)$, $(0.1, 0.3)$, 
  and $(0.1, 0.4)$, respectively.
  The SZ angular power spectra 
  for $\sigma_8 = 1.0$ and $\sigma_8 = 0.77$ in the $\Lambda$CDM model are shown
  as the thin solid and thin dashed lines, respectively.
  For references, we plot the primordial CMB temperature angular power spectrum in 
  our fiducial $\Lambda$CDM model and the ACBAR data.
  }
  \label{fig:szcl_m26.eps}
\end{figure}

Figure~\ref{fig:szcl_m26.eps} shows the SZ power spectrum in the modified gravity model. 
We take three sets of the model parameters:
$(\alpha,r_cH_0)=(0.1, 0.2)$ , $(0.1, 0.3)$, and $(0.1, 0.4)$, giving $\sigma_8=1.2$, $1.0$ and $0.9$, respectively (The initial density power spectra are the same for all of these parameter sets).
For comparison, we plot in Fig.~\ref{fig:szcl_m26.eps}
the SZ power spectra in the $\Lambda$CDM models
with $\sigma_8=0.77$ (our fiducial model) and with $\sigma_8=1.0$.
One sees that the amplitude of the peak in modified gravity is
lower than that in the $\Lambda$CDM model with the same $\sigma_8$.
This is because the actual density in a virialised halo is 
lower in modified gravity than in the $\Lambda$CDM model.

\begin{figure}
  \begin{center}
  \includegraphics[keepaspectratio=true,height=60mm]{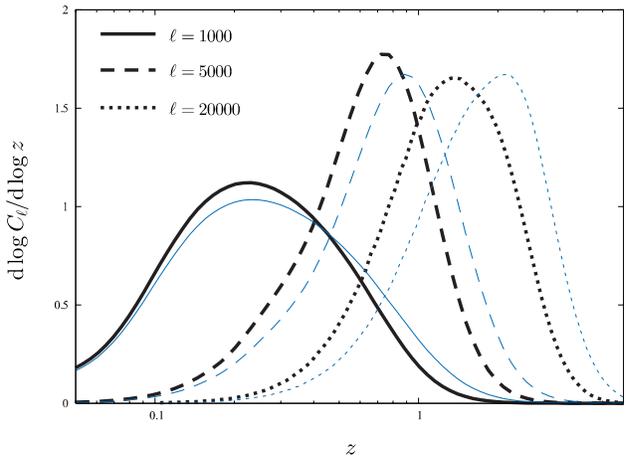}
  \end{center}
  \caption{Distribution of the redshift contribution of the SZ angular power spectrum
 for different $\ell$ modes.
 We set the modified gravity parameters $\alpha=0.1$ and $r_cH_0=0.3$.
 The solid, the dashed, and the dotted lines
 represent the distributions for $\ell=1000$, $\ell=5000$, and
$\ell=20000$, respectively.
 For comparison, we plot the distributions for the $\Lambda$CDM model
 with $\sigma_8=1.0$ as thin lines.
  }
  \label{fig:redshift_dist.eps}
\end{figure}

Moreover, modified gravity shows the damping of the SZ power spectrum on small scales 
and the amplification on large scales,
because the contribution from halos at high redshifts in modified gravity 
is smaller than that in the $\Lambda$CDM model.
The redshift contribution for different $\ell$ is shown in Fig.~\ref{fig:redshift_dist.eps}.
The growth of matter density fluctuations in the modified gravity model is rapid 
at late times compared with the growth in the $\Lambda$CDM model with the same $\sigma_8$.
As a result, the formation of halos delays,
leading to the large contribution from low redshifts and small contribution from high redshifts.

\begin{figure}
  \begin{center}
  \includegraphics[keepaspectratio=true,height=60mm]{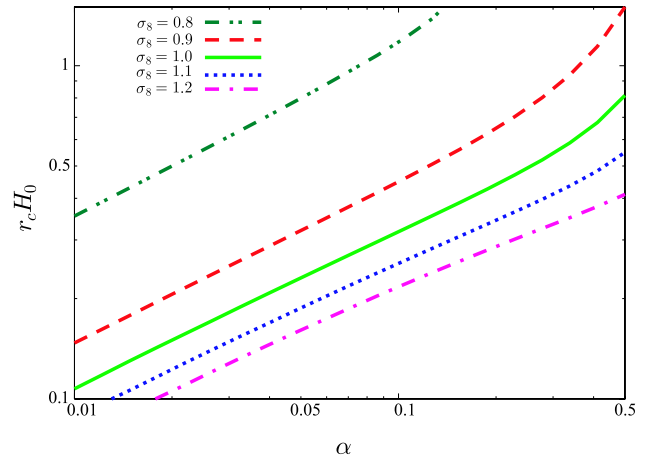}
  \end{center}
  \caption{
  The amplitude $\sigma_8$ as a function of modified gravity parameter $\alpha$ and $r_c$.
  The dashed-dotted-dotted, dashed, solid, dotted, dashed-dotted lines represent
  $\sigma_8=0.8$, $0.9$, $1.0$, $1.1$ and $1.2$.
  }
  \label{fig:a-b_m26.eps}
\end{figure}

The SZ power spectrum is constrained by the
CMB observations on small scales. 
In the $\Lambda$CDM model the ACBAR data at $\ell \sim 2500$ gives the constraint
on the SZ power spectrum in terms of $\sigma_8$: $\sigma_8 < 1$ \citep{acbar-kuo-2007}.
The peak amplitude of the SZ power spectrum 
is lower
in modified gravity than in the $\Lambda$CDM model with the same $\sigma_8$.
This result then yields
the constraint from the ACBAR data $\sigma_8 < 1.2$ in modified gravity.
Figure~\ref{fig:a-b_m26.eps} shows
$\sigma_8$ as a function of the model parameters $\alpha$ and $r_cH_0$.
In the limit $r_c\to\infty$, $\sigma _8$ reduces to the value in our fiducial model,
$\sigma_8=0.77$. From the ACBAR experiment,
the parameter region below the $\sigma_8=1.2$ line in Fig.~\ref{fig:a-b_m26.eps} is excluded.



Stronger constraints can be obtained from
the small angular scale CMB measurements by the QUaD experiment
\citep{quad-2009}.
The QUaD telescope measured the CMB temperature anisotropy in the multipole range 
$2000<\ell <3000$ at 150 GHz.
The QUaD team reported no strong evidence of the SZ effect and the result
is consistent with the WMAP data which yields $\sigma_8 \sim 0.8$.
Therefore, using the result of the QUaD experiment to constrain $\sigma_8$,
the modified gravity parameters are more strongly restricted than
in the case of the ACBAR data.
The constraint from the QUaD data gives $\sigma_8 \sim 0.9$ in modified gravity.

\section{Conclusions}\label{Sec:Conclusions}

In this paper, we have explored observational consequences
of structure formation in modified gravity.
The model considered is a phenomenological extension of
the DGP braneworld, and modification to the standard $\Lambda$CDM model is
characterized by the additional term $\pm H^{2\alpha}/r_c^{2(1-\alpha)}$
in the Friedmann equation.
In the case of $\alpha=1/2$, the term arises from the five-dimensional effect, but
in this work $\alpha$ and $r_c$ were assumed to be free parameters.

First, we have studied
the spherical collapse model of
nonlinear structure formation in modified gravity.
It was found that change in the growth function is relatively large, but
the linear density contrast for spherical collapse undergoes very small modification.
For the self-accelerating branch, the growth of perturbations is suppressed compared to
the $\Lambda$CDM model,
which confirms qualitatively the result of~\cite{BK}.
For the normal branch,
the growth of perturbations is enhanced compared to the $\Lambda$CDM model.

Focusing on the normal branch, we then investigated
the effect of modified gravity on the SZ angular power spectrum.
The enhanced growth function in the normal branch results in
the amplification of $\sigma_8$.
However, the modification to the SZ power
spectrum is rather nontrivial.
The peak amplitude
of the SZ angular power spectrum is lower in modified gravity
than in the $\Lambda$CDM model with the same $\sigma_8$, because halos are virialised
at lower density in modified gravity than in the $\Lambda$CDM model.
In addition to this, modified gravity shows the damping of the SZ power spectrum on small scales.


We confronted the modified SZ spectrum with CMB observations on small scales.
Observational constraints can be read off from Fig.~\ref{fig:a-b_m26.eps}.
The ACBAR experiment gives the constraint on
the SZ power spectrum which translates to $\sigma_8=1.2$ in modified gravity.
From this we put constraints to the parameters of the modified gravity model:
for example, we have
 $r_cH_0  > 0.4 $ for $\alpha=0.5$ and $r_cH_0 > 0.2 $ for $\alpha=0.1$.
The QUaD data gives severer constraints on the parameters: $\sigma_8=0.9$, leading to 
$r_cH_0 > 0.4 $ for $\alpha=0.1$.



\section*{Acknowledgments}

We would like to thank Bj{\"o}rn Malte Sch{\"a}fer and Kazuya Koyama
for helpful correspondences and
Naoshi Sugiyama for useful comments.
We also would like to thank an anonymous referee for useful comments to improve the paper.
TK is supported by the JSPS under Contact No.~19-4199.

\appendix
\section{Comparison with different approaches}

\begin{figure}
  \begin{center}
    \includegraphics[keepaspectratio=true,height=50mm]{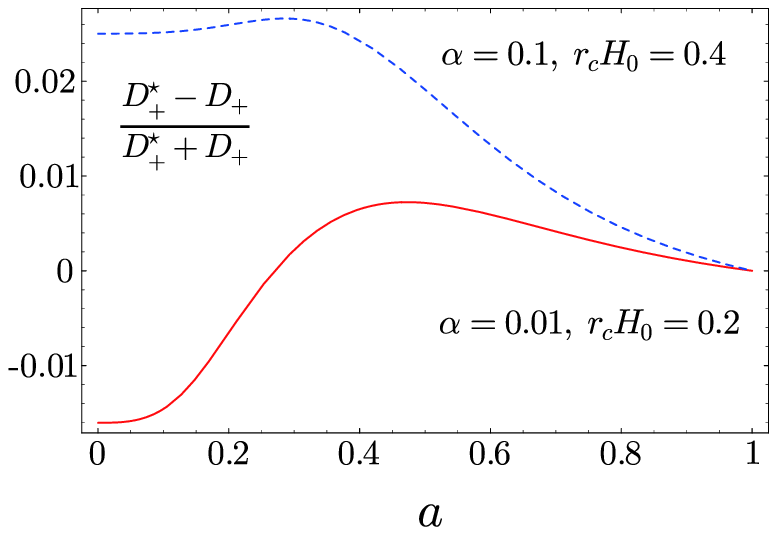}
  \end{center}
  \caption{Comparison with the growth functions obtained by the
  different approaches.}%
  \label{fig:comp.eps}
\end{figure}

In the phenomenologically extended version of the DGP model,
different assumptions can be made in computing the growth of perturbations.
Instead of using Birkhoff's theorem, we may assume
that the behaviour of perturbations is governed by a scalar-tensor theory.
Following~\cite{Koyama^alpha}, we now compare the linear evolution
of perturbations obtained by these two approaches.
If one does not assume Birkhoff's theorem
but employs a scalar-tensor theory as an effective theory
for perturbations,
the evolution of $\delta$ will be described by~\citep{Koyama^alpha}
\begin{eqnarray}
\ddot\delta+2H\dot\delta = 4\pi G\left(1+\frac{1}{3\beta}\right)\rho\delta,
\end{eqnarray}
where
\begin{eqnarray}
\beta:=1+\frac{(r_c H)^{2(1-\alpha)}}{\alpha}\left[
1+\frac{2}{3}\frac{(1-\alpha)\dot H}{H^2}
\right].
\end{eqnarray}
Note that this is the expression for the normal branch.
We denote by
$D_+^\star (a)$
the growth function calculated from this equation.
The difference between $D_+^\star$ and $D_+$ obtained in the main text,
$(D_+^\star-D_+)/(D_+^\star+D_+)$, is presented in Fig.~\ref{fig:comp.eps}.
As can be seen, there is only a few percent difference between the two approaches
concerning the linear growth of perturbations.
Developing nonlinear theory with the violation of Birkhoff's theorem
taken into account is beyond the scope of the present paper.

\end{document}